\documentclass[12pt]{article}

\catcode`\@=11

\def\section{\@startsection{section}{1}{\z@}{3.5ex plus 1ex minus
 .2ex}{2.3ex plus .2ex}{\bf}}

\def\thesubsection{\arabic{section}.\arabic{subsection}}
\renewcommand{\subsection}[1]{\addtocounter{subsection}{1}
\vspace{2.5mm}\par\noindent {\it \thesubsection . #1}\par
 \vspace{0.5mm} }
\catcode`\@=12
\newfont{\mbm}{msbm10 scaled\magstep1}

\newcommand{\nc}{\newcommand}

\nc{\be}{\begin{equation}}
\nc{\ee}{\end{equation}}
\nc{\bea}{\begin{eqnarray}}
\nc{\eea}{\end{eqnarray}}

\begin{document}
\begin{titlepage}

\rightline{ROM2F-2001/02}
\rightline{{hep-th/0101085}}
\vskip 3cm
\centerline{{\large\bf  Type-I vacua from non-geometric orbifolds}}
\vskip 1.5cm
\centerline{Gianfranco Pradisi}
\vskip 1.2cm
\centerline{\it Dipartimento di Fisica, Universit\'a di Roma
``Tor Vergata'' e INFN, Sezione di Roma 2}
\centerline{\it Via della Ricerca Scientifica 1, I-00133 Rome}
\vskip  2.0cm

\begin{abstract}
{Some examples of Type-I vacua related to  non geometric orbifolds 
are shown.  In particular, the open descendants of the diagonal $Z_3$ 
orbifold are compared with the geometric ones. Although not chiral, these 
models exhibit some interesting properties, like twisted sectors in the 
open-string spectra and the presence of ``quantized'' geometric moduli, 
a key ingredient to ensure their perturbative consistency and to explain 
the rank reduction of their Chan-Paton groups.}
\end{abstract}
\vskip 36pt
\begin{center}
{\it 
Contribution to the Proceedings of IX Marcel Grossman Meeting\vskip 18pt
 Rome, ITALY, July 2-8, 2000}
\end{center}

\end{titlepage}
\section{Introduction}
With the advent of superstring dualities\cite{sen1}, all the five known superstring theories (Type I, Type II and Heterotic) have been recognized as differentperturbative  asymptotics of the underlying (but still unknown) eleven dimensional M-Theory\cite{wit1}, whose low energy limit is the eleven dimensional supergravity.  Despite this unification, the resulting web of lower dimensional models is quite complicated, and a systematic classification of all possible vacua is still lacking.  Moreover, several ways are today conceivable to embed the Standard Model within this scheme: besides the heterotic susy-GUT scenario\cite{die1}, where the string scale is forced to be close to the Planck scale, there is a suggestive ``brane-world'' scenario\cite{Dudrev}, in which the string scale could be lowered everywhere between the Planck scale and a few TeV. These models, based on Type I vacua\cite{gs,car} (possibly supplemented with additional brane-antibrane pairs) contain gauge fields living only on a four-dimensional defect, while gravitational interactions propagate in the ten dimensional bulk.  As a result, due to the smallness of the Newton constant, some of the extra-dimensions are allowed to be large, even of millimeter size\cite{led}.  Thus, fundamental issues like the gauge hierarchy problem and the breaking of supersymmetry should be reconsidered.  In this paper, I will review some examples of Type I vacua built as open descendants\cite{prsa} of closed Type II orbifold models.  In particular, I will describe the Type I descendant of the $Z_3$ geometric orbifold\cite{chir}, the first four-dimensional model exhibiting three generations of chiral matter with gauge group $SO(8) \times U(12)$, the simplest non-geometric Type I descendant of the $Z_3$-orbifold, namely the diagonal model,\cite{diag} and some non-geometric asymmetric orbifold examples\cite{bimopra}, that exhibit interesting features like discrete torsion and ``brane supersymmetry breaking''\cite{bsb1}.

\section{The Type I geometric $Z_3$-orbifold vacua in $D=4$}

Orbifold compactifications\cite{orb1} were discovered long ago as the simplest examples of string compactifications that reduce the number of supersymmetries and allow fermion chirality.  The Type IIB $Z$ orbifold is a four dimensional model obtained modding out a product of three two-tori by the geometric action of the $Z_3$ group $X \sim \omega^k X$, where $\omega = e^{2 i \pi/3}$.  The resulting (closed) partition function corresponds to (an irrational deformation of) the so called charge-conjugation modular invariant \footnote{the notation follows ref. (\cite{diag})}
\bea
{\cal T_{\it{c}}} &\!=&\! {1 \over 3} \, {\bigg[} \ \rho_{00} \,
\bar\rho_{00} \  {\prod_{i=1}^{3} \Lambda_{i}}  + \rho_{01}\bar\rho_{02} + 
\rho_{02}\bar\rho_{01} + \ 27 \ ( \ \rho_{10}\bar\rho_{20} + \rho_{11}\bar\rho_{22} 
+ \rho_{12}\bar\rho_{21} \ )  \nonumber \\  
&\!+&\! 27 \ ( \ \rho_{20}\bar\rho_{10} + \rho_{21}\bar\rho_{12}
+  \rho_{22}\bar\rho_{11} \ ) \ {\bigg]} \qquad,
\label{dd6modinv}
\eea
where $\Lambda_{i}$ are the Narain lattice sums on the three hexagonal lattices defining the tori, with vanishing values of the NS-NS antisymmetric tensor $B$.  The coefficients in front of the twisted sector amplitudes are due to the fixed points.  The projection preserves $N=2$ supersymmetry and produces a spectrum whose untwisted part contains the $N=2$ supergravity multiplet, the universal dilaton hypermultiplet and $9$ additional hypermultiplets.  There are also $27$ hypermultiplets coming from twisted sectors, that account for a compactification on a Calabi Yau (CY) threefold with Hodge numbers $h_{1,1}=36$ and $h_{1,2}=0$. Notice that the geometric action corresponds to a ``non diagonal'' form of the partition function, reflected in the sesquilinear combination of chiral supersymmetric conformal blocks, each paired with the complex conjugate in (\ref{dd6modinv}).
To build the Type I descendants\cite{chir}, one starts by modding out the closed sector with the (closed) world-sheet parity operator $\Omega$ or, equivalently, by halving the torus contribution and adding the Klein-bottle amplitude.  The resulting (unoriented) spectrum comprises the $N=1$ supergravity multiplet together with $9$ linear multiplets from the untwisted sector and $27$ linear multiplets from the twisted ones.  The projection produces uncancelled tadpoles or, equivalently, introduces $O9$-planes carrying negative RR charge requiring the inclusion of open and unoriented amplitudes or of $32$ D9-branes.  Indeed,  $Z_3$ does not contain order two elements and only open strings with Neumann boundary conditions on the two ends are allowed.  Consistency conditions (i.e. tadpole cancellations) from twisted and untwisted transverse sectors fix the gauge Chan Paton (CP) group to be $SO(8) \times SU(12) \times U(1)$, with three generations of chiral multiplets in $(\bf8,\bf{\bar{12}})_{-1}$ and $(\bf1,\bf{66})_{2}$. The $U(1)$ factor is anomalous and requires a modified version of the GSS mechanism,\cite{gs,gss} in which the dilaton plays no role and the anomaly cancellation is due to a coupling of the gauge fields with a combination of twisted closed moduli\cite{tad}.  A non-vanishing quantized $B$\cite{tor} can also be introduced with a consequent rank reduction of the CP group.  The resulting models can be found in ref. (\cite{carlo}).

\section{The Type I diagonal  $Z_3$-orbifold vacua in $D=4$}

The diagonal partition function corresponds to an orbifold by a combination of the geometric action and an involution of order two that conjugates the eigenvalues of only the right-moving coordinates.  Equivalently, it can be obtained by combining the geometric action with a T-duality along three of the six real coordinates.  Since T-duality exchanges Type IIB with Type IIA, the resulting Type I vacua can also be regarded as geometric orientifolds of the Type IIA superstring, also known in the literature as Type I' models.  One can perform the orientifold projection in two equivalent ways:  as the combination of $\Omega$ and an involution $\cal I$ of order two on the geometric modular invariant\cite{bgk} or as the $\Omega$ projection on the diagonal modular invariant\cite{diag}.  The last procedure has been successfully used at rational points in several examples, like Gepner models\cite{gep} and free-fermionic constructions\cite{bs}.  In irrational models, however, there is a subtlety connected to the projection of the Narain lattice sums.  Indeed, due to the T-dualities, the projected lattice is a skew slice containing both momenta and windings, while the left-right symmetry quantizes the off diagonal components of the background metric rather than of $B$, that is now a continuous modulus\cite{cb}.  As a result, even if one chooses a vanishing $B$, the rank of the CP group is reduced by a factor of two for each complex compact coordinate.  The quantized geometric moduli are key ingredients of the construction: they account for the correct reflection coefficients in front of the crosscaps and enforce the complete projector in the transverse channel in such a way that only the identity character can flow along the tube at massless level, a feature very familiar from the rational diagonal models.  There is another subtlety connected to the diagonal action on the lattice.  From a geometrical point of view T-duality can be regarded as a reflection with respect to some symmetry axis that leaves invariant only a subset of the $27$ fixed points.  For $Z_3$, it is easy to see that two reflections ${\cal{I}}_1$ and  ${\cal{I}}_2$ are compatible with the geometric action , the former preserving only the origin of the lattice and the latter preserving all the three fixed points in each of the compact directions.  The two reflections correspond to different resolutions of the ``fixed-point ambiguities''\cite{Gep1} related to identical closed twist fields sitting at the orbifold singularities.  Combinations of the two actions thus yield four classes of models containing $1$, $3$, $9$ or $27$ chiral twist fields surviving the unoriented projection.  The massless spectra of the resulting models are reported in Table 1. 
\begin{table}
\begin{center}
\begin{tabular}{|c|c|c||c|c||c|c|} 
\hline
\, & closed & closed & open & open & open & open \\
\hline 
$n$ & {\bf{C}} & {\bf{V}} & CP & {\bf{C}} & CP & {\bf{C}} \\
\hline
\hline
$1$ & 24 & 13 & $SO(4)$ & 3$({\bf{6}})$+1$({\bf{10}})$ & $Sp(4)$ & 4$({\bf{10}})$ \\
\hline
$3$ & 25 & 12 & $SO(4)$ & 3$({\bf{6}})$+3$({\bf{10}})$ & $Sp(4)$ & 6$({\bf{10}})$ \\
\hline
$9$ & 28 & 9 & $SO(4)$ & 3$({\bf{6}})$+9$({\bf{10}})$ & $Sp(4)$ & 12$({\bf{10}})$ \\
\hline
$27$ & 37 & 0 & $SO(4)$ & 3$({\bf{6}})$+27$({\bf{10}})$ & $Sp(4)$ & 30$({\bf{10}})$ \\
\hline
\end{tabular}
\end{center}
\caption{Massless spectra of $D=4$ models.}
\label{tab4}
\end{table}
Notice that the closed sector corresponds to a compactification on a CY threefold with Hodge number $h_{1,1}=0$ and $h_{1,2}=36$.  Not surprisingly, in this context T-duality is thus equivalent to mirror symmetry.  Twisted open strings are allowed in the annulus and M\"obius amplitudes.  Thus, a single type of CP charge is present, associated to D6 branes at angle \cite{blum}, the natural objects after three T-dualities.  The size of the CP group is fixed by a single tadpole condition and receives contributions solely from the extended coordinates.  This reflect itself in the mentioned rank reduction from 16 to 2.  A discrete Wilson line is also allowed, that gives rise to orthogonal or symplectic gauge groups.  Amazingly, the construction seems to be consistent for all odd-integer values of $n$ between $1$ and $27$, provided an involution on the lattice preserving exactly $n$ out of the $27$ fixed-points exists.

\section{Type I vacua from asymmetric orbifolds}

Interesting classes of non-geometric Type I vacua can also be constructed starting from asymmetric orbifold compactifications\cite{nsv}, obtained quotienting a torus compactification by a group acting differently on the left and right modes.  For this action to be allowed, the lattice must admit a chiral isometry, an option available only at rational points.   Moreover, there is generically the possibility of turning on some discrete torsion connecting, in this context, models with different amount of supersymmetry.  In ref. (\cite{bimopra}) two examples are reported, a $Z_2 \times Z_2$ asymmetric orbifold in six dimensions and a $Z_3 \times Z_3$ asymmetric orbifolds both in six and in four dimensions.  The resulting massless spectra are summarized in table 2, where $\epsilon$ is the discrete torsion and $\sum_{i=1}^{4} n_i=16$ in the first row. 
\begin{table}
\begin{center}
\begin{tabular}{|c|c|c|c|c|c|c|} 
\hline
D & $\epsilon$ & model& susy & closed & CP & open\\
\hline
\hline 
$6$ & 1 & $Z_2\times Z_2$ & N=(1,1) & {\bf{G}}+4{\bf{V}} & $\prod_{i= 1}^{4} Sp(n_{i})$ & - \\
\hline
$6$ & -1 & $Z_2\times Z_2$ & N=(1,0) & {\bf{G}}+14{\bf{H}}+7{\bf{T}} & $Sp(4)^4$ & 6 {\bf{H}}$({\bf{4}},{\bf{4}})$\\
\hline
\hline
$6$ & 1 & $Z_3\times Z_3$ & N=(1,1) &{\bf{G}}+4{\bf{V}}&SO(8)& - \\
\hline
$6$ & $\omega$,$\bar{\omega}$ & $Z_3\times Z_3$ & N=(1,0) & {\bf{G}}+15{\bf{H}}+7{\bf{T}} & $SO(8)$ & 4 {\bf{H}} $({\bf{28}})$\\
\hline
\hline
$4$ & 1 & $Z_3\times Z_3$ & N=2 & {\bf{G}}+2{\bf{V}} + 9 {\bf{H}} & $Sp(4)$ & 4 {\bf{H}}$({\bf{10}})$\\
\hline
$4$ & $\omega$ & $Z_3\times Z_3$ & N=1 & {\bf{G}}+25{\bf{C}} & $Sp(4)$ & 6 {\bf{H}}$({\bf{10}})$\\
\hline
$4$ & $\bar{\omega}$ & $Z_3\times Z_3$ & N=1 & {\bf{G}}+3{\bf{V}}+22{\bf{C}} & $Sp(4)$ & 12 {\bf{H}}$({\bf{10}})$\\
\hline
\end{tabular}
\end{center}
\caption{Massless spectra of asymmetric orbifold  models.}
\label{tab2}
\end{table}
Notice that in $D=6$ the models with $\epsilon=1$ correspond to compactifications on $T^4$, while the models with $\epsilon\ne 1$ correspond to compactifications on $K_3$.  In D=4, the models with $\epsilon=1$ correspond to compactifications on $K_3\times T^2$ while the models with $\epsilon\ne 1$ correspond to compactifications on two mirror CY spaces with Hodge numbers $h_{1,1}=18= h_{1,2}^{\prime}$, $h_{1,2}=6=h_{1,1}^{\prime}$.  The rank reduction is due to the presence of a quantized $B$ or, equivalently, to compactifications without vector structure.  All these models are not chiral.
Interestingly, brane supersymmetry breaking can be introduced also in this non-geometric setting.  For instance, a consistent variant of the $Z_2 \times Z_2$ asymmetric model in six dimensions with $\epsilon=-1$ can be build replacing a pair of D-branes with the corresponding anti-D-branes, while introducing a different $\Omega$ projection on the fixed-point space.  The new $\Omega$ gives rise to models with closed unoriented massless (supersymmetric) spectra containing the $N=(1,0)$ supergravity multiplet coupled to 10 hypermultiplets and 11 tensor multiplets.  In the open sector, the gauge group is $U(4)^2$ and supersymmetry is broken by open strings charged under the anti-D-brane gauge group, allowing spinors in the bifundamental representation and conjugate spinors in the symmetric representation.  The vanishing of R-R tadpoles ensures, as usual, the absence of both gauge and gravitational anomalies.  It would be interesting to extend further the analysis of non-geometric compactifications and to clarify the relation with non BPS states\cite{sen2}.  A step forward in this direction has been made in ref. (\cite{cbg}).

This work was supported in part by the EEC contract HPRN-CT-200-00122 and by the INTAS contract 9915590.


\begin{thebibliography}{99}

\bibitem{sen1}See, for instance, A.~Sen,
Nucl.\ Phys.\ Proc.\ Suppl.\ {\bf 58} (1997) 5
[hep-th/9609176].

\bibitem{wit1} E.~Witten,
Nucl.\ Phys.\ {\bf B443} (1995) 85
[hep-th/9503124];
C.~M.~Hull and P.~K.~Townsend,
Nucl.\ Phys.\ {\bf B451} (1995) 525
[hep-th/9505073].

\bibitem{die1}See, for instance, K.~R.~Dienes,
Phys.\ Rept.\ {\bf 287} (1997) 447
[hep-th/9602045].

\bibitem{Dudrev}
See, for instance, E.~Dudas,
Class.\ Quant.\ Grav.\ {\bf 17} (2000) R41
[hep-ph/0006190].

\bibitem{gs} M.~B.~Green and J.~H.~Schwarz,
Phys.\ Lett.\ {\bf B149} (1984) 117.
Phys.\ Lett.\ {\bf B151} (1985) 21.

\bibitem{car}
A.~Sagnotti,
ROM2F-87-25
{\it Talk presented at the Cargese Summer Institute on Non-Perturbative Methods in Field Theory, Cargese, France, Jul 16-30, 1987}.

\bibitem{led} See, for instance, 
I.~Antoniadis and K.~Benakli,
Int.\ J.\ Mod.\ Phys.\ {\bf A15} (2000) 4237
[hep-ph/0007226].

\bibitem{prsa}
G.~Pradisi and A.~Sagnotti,
Phys.\ Lett.\ {\bf B216} (1989) 59.

\bibitem{chir}
C.~Angelantonj, M.~Bianchi, G.~Pradisi, A.~Sagnotti and Y.~S.~Stanev,
Phys.\ Lett.\ {\bf B385} (1996) 96
[hep-th/9606169].

\bibitem{diag} G.~Pradisi,
Nucl.\ Phys.\ {\bf B575} (2000) 134
[hep-th/9912218].

\bibitem{bimopra} M.~Bianchi, J.~F.~Morales and G.~Pradisi,
Nucl.\ Phys.\ {\bf B573} (2000) 314
[hep-th/9910228].

\bibitem{bsb1}
I.~Antoniadis, E.~Dudas and A.~Sagnotti,
Phys.\ Lett.\ {\bf B464} (1999) 38
[hep-th/9908023];
G.~Aldazabal and A.~M.~Uranga,
JHEP{\bf 9910} (1999) 024
[hep-th/9908072].

\bibitem{orb1} L.~Dixon, J.~A.~Harvey, C.~Vafa and E.~Witten,
Nucl.\ Phys.\ {\bf B261} (1985) 678.

\bibitem{gss} A.~Sagnotti,
Phys.\ Lett.\ {\bf B294} (1992) 196
[hep-th/9210127].

\bibitem{tad} See, for instance, H.~P.~Nilles,
hep-th/0003152.

\bibitem{tor}
M.~Bianchi, G.~Pradisi and A.~Sagnotti,
Nucl.\ Phys.\ {\bf B376} (1992) 365.

\bibitem{carlo} C.~Angelantonj,
Nucl.\ Phys.\ {\bf B566} (2000) 126
[hep-th/9908064].

\bibitem{bgk} R.~Blumenhagen, L.~Gorlich and B.~Kors,
Nucl.\ Phys.\ {\bf B569} (2000) 209
[hep-th/9908130].

\bibitem{gep}C.~Angelantonj, M.~Bianchi, G.~Pradisi, A.~Sagnotti and Y.~S.~Stanev,
Phys.\ Lett.\ {\bf B387} (1996) 743
[hep-th/9607229].

\bibitem{bs}
M.~Bianchi and A.~Sagnotti,
Phys.\ Lett.\ {\bf B247} (1990) 517,
Nucl.\ Phys.\ {\bf B361} (1991) 519.

\bibitem{cb} C.~Angelantonj and R.~Blumenhagen,
Phys.\ Lett.\ {\bf B473} (2000) 86
[hep-th/9911190].

\bibitem{Gep1}
D.~Gepner,
Phys.\ Lett.\ {\bf B222} (1989) 207;
J.~Fuchs, B.~Schellekens and C.~Schweigert,
hep-th/9612093,
Nucl.\ Phys.\ {\bf B473} (1996) 323
[hep-th/9601078],
Nucl.\ Phys.\ {\bf B461} (1996) 371
[hep-th/9509105].

\bibitem{blum}
R.~Blumenhagen, L.~Gorlich and B.~Kors,
JHEP{\bf 0001} (2000) 040
[hep-th/9912204].

\bibitem{nsv}
K.~S.~Narain, M.~H.~Sarmadi and C.~Vafa,
Nucl.\ Phys.\ {\bf B288} (1987) 551.

\bibitem{sen2}
A.~Sen,
hep-th/9904207.

\bibitem{cbg}
C.~Angelantonj, R.~Blumenhagen and M.~R.~Gaberdiel,
Nucl.\ Phys.\ {\bf B589} (2000) 545
[hep-th/0006033].
\end{thebibliography}
\end{document}